\documentclass[twocolumn,english,aps,prl,superscriptaddress,preprintnumbers]{revtex4}
\usepackage[T1]{fontenc}
\usepackage[latin9]{inputenc}
\usepackage{amsmath}
\usepackage{amssymb}
\usepackage{graphicx}

\usepackage{hyperref}

\hypersetup{
    colorlinks=true,
    linkcolor=black,
    citecolor=black,
    filecolor=black,
    urlcolor=black,
}

\makeatletter

\usepackage{amstext}

\usepackage{babel}

\makeatother

\begin{document}

\title{Efficient mode conversion in an optical nanoantenna mediated by quantum emitters}

\author{J. Straubel}
\affiliation{Institute of Theoretical Solid State Physics, Karlsruhe Institute of Technology, 76131 Karlsruhe, Germany}
\author{R. Filter}
\affiliation{Institute of Condensed Matter Theory and Solid State Optics, Abbe Center of Photonics, Friedrich-Schiller-Universit\"{a}t Jena, 07743 Jena, Germany}
\author{C. Rockstuhl}
\affiliation{Institute of Theoretical Solid State Physics, Karlsruhe Institute of Technology, 76131 Karlsruhe, Germany}
\affiliation{Institute of Nanotechnology, Karlsruhe Institute of Technology, 76021 Karlsruhe, Germany}
\author{K. S\l owik}
\email{karolina@fizyka.umk.pl}
\affiliation{Institute of Theoretical Solid State Physics, Karlsruhe Institute of Technology, 76131 Karlsruhe, Germany}
\affiliation{Instytut Fizyki, Uniwersytet Miko\l{}aja Kopernika, 87-100 Toru\'{n}, Poland}

\begin{abstract}
Converting signals at low intensities between different electromagnetic modes is an asset for future information technologies. In general, slightly asymmetric optical nanoantennas enable the coupling between bright and dark modes that they sustain. However, the conversion efficiency might be very low. Here, we show that the additional incorporation of a quantum emitter allows to tremendously enhance this efficiency. The enhanced local density of states cycles the quantum emitter between its upper and lower level at an extremely hight rate; hence converting the energy very efficient. The process is robust with respect to possible experimental tolerances and adds a new ingredient to be exploited while studying and applying coupling phenomena in optical nanosystems.
\end{abstract}

\maketitle


The ability of plasmonic nanoantennas to control the spatial and the spectral distribution of electromagnetic fields attracted tremendous interests \cite{Biagioni2012,Alu2008,Liu2011,Maksymov2012}. Steering, operating, and converting information in different modes of the electromagnetic field is an important ability for a future highly integrated information architecture. Of particular importance has been the understanding of coupling mechanisms among different modes sustained by nanoantennas. These modes can be either bright or dark, depending on whether free space radiation from a specific input mode can excite the mode or not. The coupling \cite{Han2015} between bright and dark modes is usually enabled by a gentle break in the symmetry \cite{Fedotov2007, Zhang2013} of the nanoantenna. Whereas coupling in general might be possible, its efficiency is very often weak. 

To mitigate this problem, we suggest to incorporate a quantum emitter into the asymmetric nanoantenna. The properties of these quantum emitters can be modified by the optical environment, giving rise to huge Purcell enhancement \cite{Anger2006, Mertens2007, Curto2010, Bujak2011, Akselrod}. Quantum emitters have also been proposed and/or demonstrated to influence statistical properties of emitted light, leading to emission of single photons \cite{Schietinger2009,Esteban2010, Busson2012, Filter2014}, or entangled pairs \cite{Maksymov}. In our application, the quantum emitter experiences an enhanced local density of states that drives the transition between its upper and lower level at a very high rate. The fast cycling of the quantum emitter allows the efficient conversion between the different modes. 

The starting point of our analysis is a bimodal nanoantenna \cite{Dopf2015}. 
The nanoantenna consists of two slightly asymmetric nanorods separated by a small gap. 
This provides the necessary asymmetry to excite, in general, with a plane wave at normal incidence all modes in the nanoantenna 
and not just those modes with an even number of nodal lines in the current that are usually considered as bright \cite{Dorfmuller2010}. 
However, as we will see, this asymmetry is too minor to excite modes with an odd number of nodal lines in the current that are usually considered as dark \cite{Hentschel2010, Hentschel2011, Liu2012} to a notable extent. The scattering strength of the perturbation is too weak. Only the incorporation of the quantum emitter allows for a notable conversion. In our analysis of the hybrid system, that bases on the time evolution of the density matrix of the hybrid system, we only require the elementary modes of the nanoantenna to be characterized by single Lorentzian resonances \cite{Truegler2008,Waks2010,Savasta2010,Evangelou2011,Slowik2013,Esteban2014,Schmidt2015}. We require that the consideration can be restricted to two modes, a requirement that is met by design. 
Specifically, the system is optimized such that the emission frequency of a nitrogen-vacancy (NV) center in diamond is between the resonance frequencies of the bright and the dark mode of the nanoantenna, respectively. 

\begin{figure}[t]
\centering
\includegraphics[width=8.6cm,keepaspectratio]{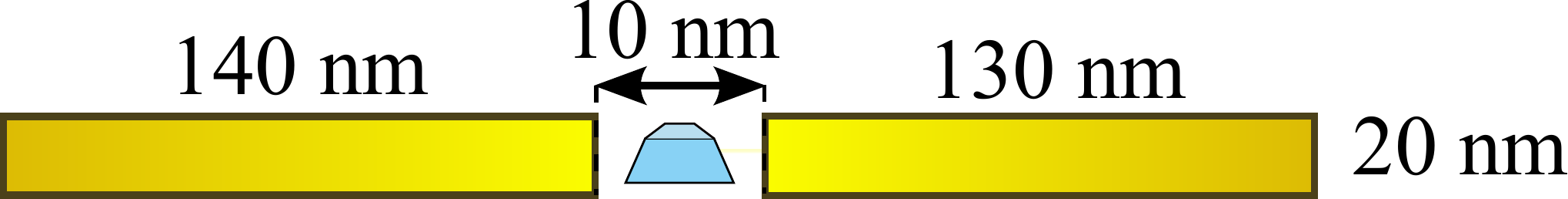}
\caption{Scheme of the hybrid structure. A nanoantenna is made from two nanorods separated by a small gap. A quantum emitter, here an NV center, is positioned in the gap.}
\label{fig:schematic}
\end{figure}

The proposed nanoantenna consists of two nanorods made from gold \cite{Palik}, with lengths of $140$ nm and $130$ nm, respectively. Both have a square cross section of $20$ $\times$ $20$ nm$^2$ and are aligned parallel to their symmetry axis, featuring a gap of $10$ nm in the center. We assume the transition of the NV center to be electric dipolar. As indicated in Fig.~\ref{fig:schematic}, the NV center is placed inside the gap of the nanoantenna. The NV center's transition dipole moment is oriented parallel to the connecting line of the nanorods. The entire structure is embedded in a glass matrix with a relative permittivity $\varepsilon = 2.25$.

\begin{figure}[tp]
\centering
\includegraphics[width=8.7cm,keepaspectratio]{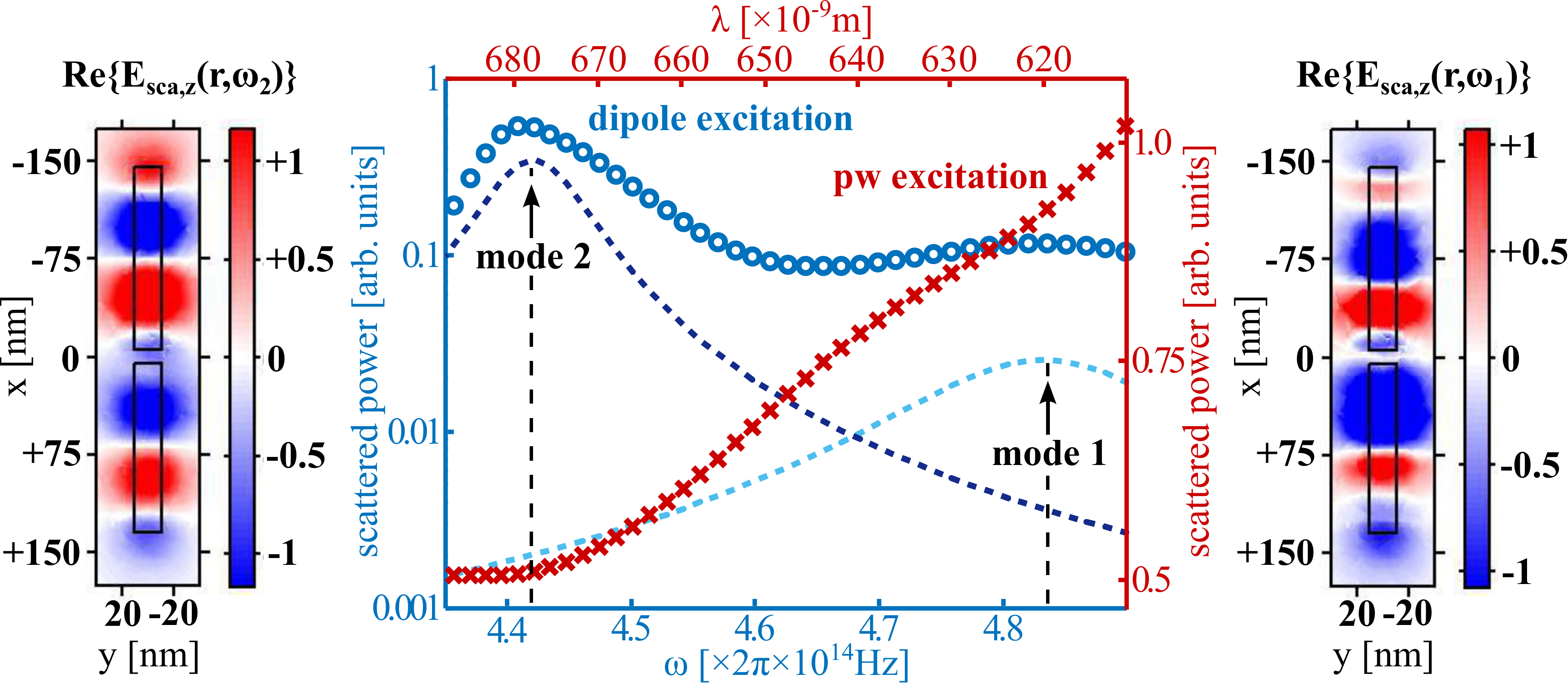}
\caption{Spectra of the scattered power by the nanoantenna for electric dipole excitation (blue circles) and plane wave illumination. Additionally shown (dashed lines): scattered power contributions of the two distinct nanoantenna modes, whose field resonances are spectrally assumed to be of Lorentzian shape. On either side: scattered field pattern of each mode on resonance with visible nodal lines. Displayed is the real part of the scattered electric field component normal to the xy-plane $20$ nm above the symmetry axis in arbitrary units for both modes each on resonance upon excitation with the dipole.}
\label{fig:spectra}
\end{figure}

The spectral response of the nanoantenna is calculated with COMSOL \textsuperscript{\textregistered}. 
Figure~\ref{fig:spectra} shows the power scattered by the nanoantenna upon plane wave (red crosses) 
and with an electric dipole (blue circles) excitation, respectively. 
The electric field of the plane wave is polarized linearly along the symmetry axis of the nanorods 
and propagation is parallel to the surface normal of one long side of the nanorods. 
Spatial position of the dipole and its orientation correspond to the spatial position and orientation of the NV center. 
For the dipole excitation, two distinct modes are discernable. 
In the following the mode resonant at $\omega_{1} = 2\pi \times 4.84\times 10^{14}$~Hz is called mode $1$. 
The mode resonant at $\omega_{2} = 2\pi \times 4.42\times 10^{14}$~Hz is called mode $2$. 
In contrast, for the plane wave excitation only mode 1 is excited. 
Please note that Figure~\ref{fig:spectra} shows the combined resonance of mode 1 and multiple other modes of higher resonance frequency, 
while the scattered power exhibits a local minimum at the resonance frequency of mode 2. 
The modes with higher resonance frequency than mode 1 are far detuned from the NV center emission frequency, and contribute to an approximately flat background. Their influence on the conversion process between modes 1 and 2 will later be included through their total Purcell factor.
Obviously, mode 1 corresponds to the bright mode whereas mode 2 - to the dark mode. 
Eventually, they correspond to modes of a nanowire antenna with consecutive mode numbers. 
The mode number is related to the number of nodal lines in the current distribution. 
Considering the two nanorods as one nanowire antenna, the modes we employ here have a mode number 6 (bright) and 5 (dark). 
To illustrate the modes, their electric field components normal to the antenna 20 nm above the structure are plotted in Fig.~\ref{fig:schematic}. 
Please note, if the mode has $N$ nodal lines in the current, it has $N+1$ nodal lines in this specific field component. 
The combination of these two modes meets the requirements for the conversion mediated by the two-level quantum system. 
Both resonances overlap spectrally, while each can be regarded as being of isolated Lorentzian shape - 
omitting the necessity of including any direct intermodal coupling in the theoretical description \cite{Gallinet2011}. 
Please note, the consideration as uncoupled is only possible because the conversion efficiency in the asymmetric optical nanoantenna is such weak, 
i.e. it can be neglected. For a plane wave excitation only one mode can be excited.    

To feed such classical electrodynamic calculations into a quantized model, we rely on a scattered field methodology. Here, we determine the time-averaged scattered and absorbed power of the nanoantenna, $\mathrm{P}_{\mathrm{sca}}$ and $\mathrm{P}_{\mathrm{abs}}$ respectively, according to \cite{Slowik2013}:
\begin{eqnarray}
\mathrm{P}_{\mathrm{sca}} &=& \int \left\langle \mathbf{E}_{\mathrm{sca}}\left(\mathbf{r},\omega\right) \times \mathbf{H}_{\mathrm{sca}}\left(\mathbf{r},\omega\right) \right\rangle d\mathbf{A}, \nonumber\\
\mathrm{P}_{\mathrm{abs}} &=& \int \left\langle \mathbf{J}_\mathrm{ind}\left(\mathbf{r},\omega\right) \cdot \mathbf{E}_\mathrm{ind}\left(\mathbf{r},\omega\right)\right\rangle dV,
\end{eqnarray}
with the electric and magnetic scattered fields given by $\mathbf{E}_{\mathrm{sca}}\left(\mathbf{r},\omega\right)$, and $\mathbf{H}_{\mathrm{sca}}\left(\mathbf{r},\omega\right)$. The symbol  $\mathbf{E}_{\mathrm{ind}}\left(\mathbf{r},\omega\right)$ stands for the electric field induced in the nanoantenna, with the corresponding current density $\mathbf{J}_\mathrm{ind}\left(\mathbf{r},\omega\right)$. 

Please note, there appears a third mode in the spectrum for plane wave illumination at higher frequencies. Since the plane wave is only required to drive at the conversion frequency, the third mode can be neglected since it is spectrally detuned. 

Up to this point, we have demonstrated that the proposed nanoantenna complies with the above mentioned characteristics for the desired mode conversion. Before we discuss the mode conversion process, we briefly recall the cavity-QED framework used to model the process.

Since the conversion is mediated by a single two-level quantum emitter, it happens at the single-photon level. 
Hence, we will describe the two nanoantenna resonances as quantum-mechanical modes \cite{Waks2010}. 
These modes are represented by annihilation operators $a_\mathrm{j}$, with $\mathrm{j}=1,2$, of eigenfrequencies $\omega_\mathrm{j}$. 
The modes are coupled to a two-level quantum emitter, with a ground state $|g\rangle$ and excited state $|e\rangle$, 
separated by transition frequency $\omega_\mathrm{qe}$. 
We assume that $\omega_\mathrm{qe}$ lies within the frequency range in which the modes overlap. 
The quantum emitter is coupled to each of the modes with a strength $\kappa_\mathrm{j}$. 
It reflects the Purcell emission enhancement of the quantum emitter due to coupling to the mode $\mathrm{j}$. 
Mode $1$ of the investigated nanoantenna is directly driven by a laser field of frequency $\omega_\mathrm{L}$, 
with a strength expressed in terms of Rabi frequency $\Omega$. 

The Hamiltonian of the system is given in a frame rotating with the frequency of the driving field: 
\begin{eqnarray}
\mathcal{H}/\hbar & = &  \sum_{\mathrm{j}=1,2}\left(\omega_\mathrm{j}-\omega_\mathrm{L}\right)a_\mathrm{j}^\dagger a_\mathrm{j} + \left(\omega_\mathrm{qe}-\omega_\mathrm{L}\right) \sigma_+\sigma_- \nonumber \\
& & + \sum_{\mathrm{j}=1,2}\kappa_\mathrm{j}\left(\sigma_+ a_\mathrm{j}+a_\mathrm{j}^\dagger \sigma_- \right) + \Omega\left( a_1^\dagger + a_1\right),
\end{eqnarray}
where $\sigma_-=|g\rangle \langle e|$ and $\sigma_+=\sigma_-^\dagger$, are the flip operators of the two-level system. 

Energy dissipation and quantum-state dephasing are included in our model in several Lindblad terms that correspond to various loss channels:
$\mathcal{L}_c\left(\gamma \right)\rho (t) = \gamma\left\lbrace c \rho(t) c^\dagger - \frac{1}{2}\left[c^\dagger c \rho (t)+\rho (t) c^\dagger c \right]\right\rbrace$,
where $\rho(t)$ represents the density matrix of the combined system of nanoantenna modes and the quantum emitter, 
$c$ is the operator responsible for particular loss channel, and $\gamma$ is the corresponding loss rate. 

The dominant loss mechanisms are photon scattering with rates $\Gamma^\mathrm{scat}_\mathrm{j}$ 
and photon absorption with rates $\Gamma^\mathrm{abs}_\mathrm{j}$, for each mode $\mathrm{j}$. 
These channels are represented by operators $\mathcal{L}_{a_\mathrm{j}}\left(\Gamma_\mathrm{j}\right)$, 
with $\Gamma_\mathrm{j} = \Gamma^\mathrm{scat}_\mathrm{j}+\Gamma^\mathrm{abs}_\mathrm{j}$. 
They are obtained from a fit to the nanoantenna scattering and absorption spectrum, respectively.

Free-space spontaneous emission of the quantum emitter with a rate $\gamma_\mathrm{fs}$ 
is included via $\mathcal{L}_{\sigma_-}\left(\gamma_\mathrm{fs}\right)$. 
The quantum emitter may additionally be subject to dephasing, 
i.e. decay of the off-diagonal terms of its density matrix, e.g. due to environmental fluctuations. 
This process is represented by $\mathcal{L}_{\sigma_+\sigma_-}\left(\gamma_\mathrm{d}\right)$. 

The stationary Lindblad equation of the coupled system reads:
\begin{equation}
-i/\hbar \left[\mathcal{H},\rho \right] + \mathcal{L}\rho = 0,
\end{equation}
where $\mathcal{L} = \mathcal{L}_{a_\mathrm{j}}\left(\Gamma_\mathrm{j}\right) + \mathcal{L}_{\sigma_-}\left(\gamma_\mathrm{fs}\right) + \mathcal{L}_{\sigma_+\sigma_-}\left(\gamma_\mathrm{d}\right)$.
It provides the steady-state density matrix $\rho$, that allows to calculate photon emission rates discussed below. 

The scattering and absorption loss rates of the nanoantenna overcome other parameters of the investigated system by a few orders of magnitude. 
Therefore, mean photon numbers are usually close to $0$. 
For this reason, it is sufficient to truncate the Hilbert space at photon numbers as low as $10$ for mode $1$, 
which is directly driven by the laser, and only $3$ for mode $2$, that can only be excited via the quantum emitter. 
To obtain the solutions, we have implemented the model described here into the QuTiP2 quantum-optics toolbox for Python \cite{qutip2}. 

To demonstrate intermodal conversion, we exploit nanoantenna parameters corresponding to the proposed design. 
This includes the resonance frequencies $\omega_\mathrm{j}$, 
and scattering and absorption rates $\Gamma_\mathrm{1}^\mathrm{scat}=1.74\times10^{14}$ Hz, 
$\Gamma_\mathrm{1}^\mathrm{abs} = 2.65\times10^{14}$ Hz, $\Gamma_\mathrm{2}^\mathrm{scat} = 6.25\times10^{13}$ Hz, 
and $\Gamma_\mathrm{2}^\mathrm{abs} = 5.20\times10^{13}$ Hz. 
This leads to efficiencies $\eta_\mathrm{j} \equiv \Gamma_\mathrm{j}^\mathrm{scat}/\Gamma_\mathrm{j}$ 
for the two modes equal to $\eta_1 \approx 0.40$, $\eta_2 \approx 0.55$. 
Both scattering and absorption rates have been determined as full-widths at half-maxima 
of the Lorentzian line shapes fitted to the nanoantenna spectrum \cite{Koppens2011}. 
The antenna has geometrically been tuned such that the transition frequency of an NV center in diamond 
$\omega_\mathrm{qe} = 2\pi \times 4.70\times 10^{14}$ Hz belongs to the spectral region in which the two modes overlap. 
Other quantum-emitter parameters corresponding to an NV center are the following: 
dipole moment $d = 3.7\times 10^{-29}$ C$^.$m \cite{Lenef}, 
the resulting spontaneous emission rate in glass $\gamma_\mathrm{fs} = 2.2 \times 10^8$ Hz, 
and dephasing rate $\gamma_\mathrm{d} = 10^6$ Hz \cite{Maze}. 

To estimate the coupling constants $\kappa_\mathrm{j}$ of the nanoantenna modes to the quantum emitter, 
we use the cavity-QED formula for emission enhancement factor $F_\mathrm{cQED}$ \cite{Keeling,Jiamin}, 
and its classical analogon $F_\mathrm{clas}$
\begin{eqnarray}
F_\mathrm{cQED} &=& 1 + \sum_{\mathrm{j}=1,2.} \frac{\eta_\mathrm{j} \kappa_\mathrm{j}^2\Gamma_\mathrm{j}}{ \left[\left(\Gamma_\mathrm{j}/2\right)^2+\left(\omega_\mathrm{j}-\omega_\mathrm{qe}\right)^2\right]
\gamma_\mathrm{fs}}, \label{eq:FcQED}\\
F_\mathrm{clas} &=& 1+ \sum_{\mathrm{j}=1,2.}\frac{P_\mathrm{j}^\mathrm{scat}}{P_0}, \label{eq:Fclas}
\end{eqnarray}
where $P_0$ stands for the power emitted by the dipole representing the quantum emitter, and $P_\mathrm{j}^\mathrm{scat}$ corresponds to the power scattered into the nanoantenna mode $\mathrm{j}$. 
We have calculated these powers through the classical simulations, and the resulting ratios read $P_\mathrm{1}^\mathrm{scat}/P_0 \approx 1.72$ and $P_\mathrm{2}^\mathrm{scat}/P_0 \approx 5.59$. 
We estimate coupling constants by a direct comparison of the on-resonance forms of expressions (\ref{eq:FcQED}) and (\ref{eq:Fclas}) \cite{Koppens2011}.
From such comparison, we obtain $\kappa_1 \approx 3.22 \times 10^{11}$ Hz and $\kappa_2 \approx 2.53 \times 10^{11}$ Hz.

Analyzing the conversion process, we are interested in two figures of merit: 
the stationary emission rate in mode $2$, given by 
\begin{equation} \label{eq:emission_rate}
r_2=\Gamma_2^\mathrm{scat}\langle a_2^\dagger a_2\rangle, 
\end{equation}
and a conversion efficiency $\xi$ defined as the ratio of photons emitted in the dark mode, 
with respect to the total number of photons emitted in both modes 
\begin{equation}
\xi = r_2/(r_2+r_1),
\end{equation}
where $r_1$ is defined analogously to $r_2$ in Eq.~(\ref{eq:emission_rate}). 
The conversion efficiency $\xi$ may be a simple measure of probability of detection of a photon in mode $2$, 
with respect to the overall probability of detecting it in either of the two dominating modes.
Such figure of merit is important for applications in quantum control of the state of the emitted light. Please note that other emission channels may be present. 
Most importantly, the quantum emitter is subject to spontaneous emission, 
whose rate at saturation conditions is comparable to the emission rates in both modes. 
However, due to the omnidirectional emission pattern, photons generated in this process can hardly be detected \cite{Grynberg}.
Moreover, other far-detuned modes might have their minor contribution to the process. 
We emphasize that these additional emission channels will not qualitatively influence the discussed conversion process, and their quantitative contribution will not be significant for the proposed setup. An additional channel is related to the stimulated process in mode $1$ with rate proportional to $\Omega$, which is negligible for weak drives. 

\begin{figure}[htbp]
\centering
\includegraphics[width=8.4cm,keepaspectratio]{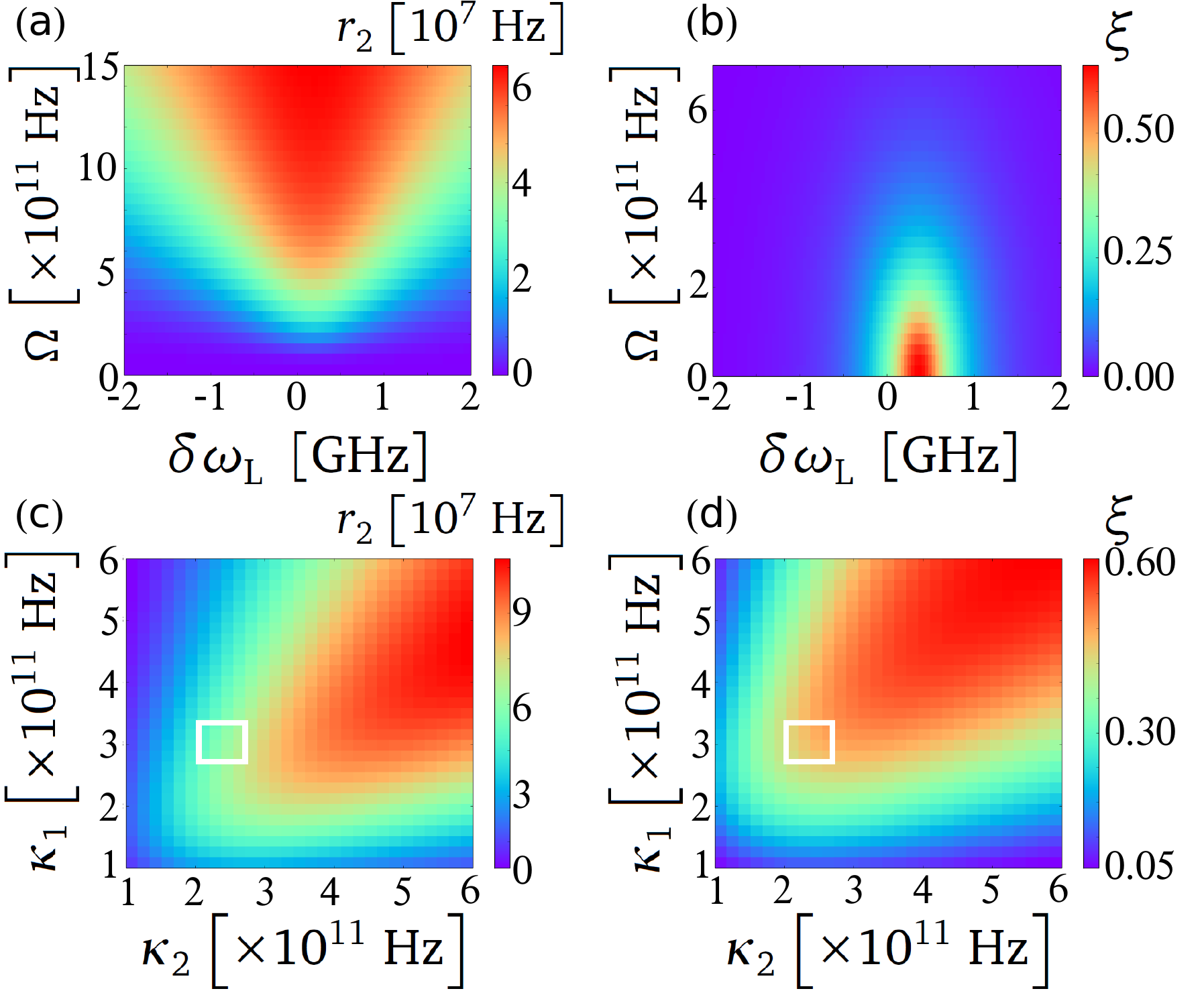}
\caption{(a,b) Steady-state values of (a) rate $r_2$ of photons emitted in mode $2$, and (b) conversion efficiency $\xi$ as functions of drive strength $\Omega$ and detuning $\delta\omega_\mathrm{L}$; (c) $r_2$ and (d) $\xi$ as functions of coupling constants $\kappa_{1,2}$ for weak drive $\Omega$ indicate robustness of the proposed scheme. Other parameters are given in the main text. The white rectangles indicate values of $\kappa$s available with the design discussed in this work, but with the quantum emitter positioned at different points within the nanoantenna gap.}
\label{fig:conversion_stationary}
\end{figure}

In Fig.~\ref{fig:conversion_stationary}(a,b) we show $r_2$ and $\xi$ as functions of experimentally tunable parameters, 
i.e. the driving field frequency detuning $\delta\omega_\mathrm{L} \equiv \omega_\mathrm{L}-\omega_\mathrm{qe}$, and strength $\Omega$. 
Naturally, the emission rate $r_2$ always grows for a fixed detuning with the driving strength, up to a point at which saturation is reached. 
Please note that the red triangular shape shown in Fig.~\ref{fig:conversion_stationary}(a) approaches the saturation regime, 
while the saturation itself is not shown here. 
For the proposed nanoantenna, driven at the NV-center transition frequency ($\delta\omega_\mathrm{L}=0$), 
the saturation takes place at $\Omega \approx 5$ THz, with $r_2 \approx 70$ MHz. 
Please note that this is already comparable to the spontaneous emission rate, which is rather large for NV centers. 

Contrary to the emission rate, the conversion efficiency $\xi$ is largest for weak drives, i.e. far from the saturation regime. 
Please note that the value of $\xi$ can exceed $1/2$, which means that a majority of photons is emitted in the plane-wave-inaccessible mode $2$. 
From Fig.~\ref{fig:conversion_stationary}(b) it follows that the conversion efficiency is considerable only if the system is driven almost resonantly. 
However, a small asymmetry is clearly visible, leading to a maximum of $\xi$ and $r_2$ at $\delta\omega_\mathrm{L} \approx 0.34$ GHz. 
This can be identified as a shift analogous to the Lamb shift in vacuum.
It arises due to the coupling of the NV center to the nanoantenna, 
and is given by $\sum_\mathrm{j} 2|\kappa_\mathrm{j}|^2\delta_\mathrm{j}/[(\Gamma_\mathrm{j}/2)^2+\delta_\mathrm{j}^2]\approx 0.336$ GHz \cite{Jiamin}, 
where $\delta_\mathrm{j} \equiv \omega_\mathrm{qe}-\omega_\mathrm{j}$. 
The Purcell broadening of the resonance with respect to the natural width $\gamma_\mathrm{fs}$ is also visible in the same Fig.~\ref{fig:conversion_stationary}(b).

In the next step, we investigate and confirm the robustness of the proposed nanoantenna geometry 
with respect to possible experimental tolerances in positioning of the quantum emitter. 
We have verified this by sweeping this position along the symmetry axis of the nanoantenna, 
with the extreme points situated as close as $2$ nm from either one of the edges of the two nanorods. 
For each location, the nanoantenna spectra are simulated individually and fitted separately. 
As expected, the resulting eigenfrequencies $\omega_\mathrm{j}$ of the modes are not modified. 
The values of coupling constants are stable and remain within the ranges of $\kappa_1 \in \left(2.73,3.36\right)\times 10^{11}$ Hz 
and $\kappa_2 \in \left(2.06, 2.68\right) \times 10^{11}~\mathrm{Hz}$. 
At the same time, the scattering and absorption loss rates remain constant to a very good approximation. 

In Fig.~\ref{fig:conversion_stationary}(c,d) we show the resulting emission rates and conversion efficiencies for a broader range of coupling constants, 
which could be obtained with modified nanoantenna geometries. 
The ranges available with the current design are marked with the white rectangle. 
An analysis of Fig.~\ref{fig:conversion_stationary}(c,d) shows the robustness of the nanoantenna design discussed in this work. 
An imprecise positioning might lead to a modification in the $r_2$ by up to $10\%$ and a negligible change of $\xi$. 
For these plots we have chosen a weak drive with a strength fixed at $\Omega = 10$ GHz, 
and always assumed a detuning $\delta\omega_\mathrm{L}$ such that the Lamb-shifted emitter resonance is hit, 
leading to optimal emission rates and conversion efficiencies. 

In conclusion, we have designed a plasmonic nanoantenna supporting two light modes in the frequency range around the NV-center transition resonance. 
The NV center, coupled to the nanoantenna, can be exploited for efficient conversion between its two eigenmodes, one of which is dark. 
Depending on the tunable properties of the field driving the system, one can switch between (1) a working mode in which high emission rates are obtained, 
exploiting the full capacity of the quantum emitter; and (2) a working mode in which a controllable and significant fraction of photons, 
even majority, is emitted in the "inaccessible" mode. 
These results are very robust against imperfect positioning of the quantum emitter. 
Possible applications of the proposed scheme include control over directionality of the emitted light, 
achieveable with tunable drive parameters, as well as quantum-state control of the emitted light. 

\textbf{Funding.} The study was supported by a research fellowship within the project 
``Enhancing Educational Potential of Nicolaus Copernicus University in the Disciplines of Mathematical and Natural Sciences'' 
(project no. POKL.04.01.01-00-081/10.) and the Karlsruhe School of Optics and Photonics (KSOP).

\end{document}